\documentclass[conference]{IEEEtran}
\IEEEoverridecommandlockouts

\usepackage{cite}
\usepackage{amsmath,amssymb,amsfonts}
\usepackage{algorithmic}
\usepackage{graphicx}
\usepackage{textcomp}
\usepackage{xcolor}
\usepackage{tikz}
\usetikzlibrary{arrows.meta,positioning}
\def\BibTeX{{\rm B\kern-.05em{\sc i\kern-.025em b}\kern-.08em
    T\kern-.1667em\lower.7ex\hbox{E}\kern-.125emX}}
\begin{document}

\title{Query Translation vs. Cross-Lingual Embeddings for Sinhala–Tamil E-Government Information Retrieval}


\author{\IEEEauthorblockN{1\textsuperscript{st} Dharshi Balasubramaniyam}
\IEEEauthorblockA{\textit{University of Kelaniya} \\
Sri Lanka\\
dharshib.8@gmail.com}
\and
\IEEEauthorblockN{2\textsuperscript{nd} Tiroshan Madushanka}
\IEEEauthorblockA{\textit{University of Kelaniya} \\
Sri Lanka\\
tiroshanm@kln.ac.lk}
}

\maketitle

\begin{abstract}
This paper presents a comparative evaluation of cross-lingual information retrieval (CLIR) methods for retrieving English government information using Sinhala and Tamil queries. Two CLIR paradigms are investigated: Query Translation (QT), employing Google Translate, NLLB, and mBART50, and Cross-Lingual Embeddings (CLE), using LaBSE, multilingual E5, and BGE-M3, with monolingual English retrieval as the baseline. Experiments are conducted on a human-verified benchmark comprising 500 Sinhala, Tamil, and English question-answer pairs derived from 1,699 segmented contexts from Sri Lanka's Government Information Center (GIC). Retrieval performance is evaluated using Recall@k (k = 1, 3, 5, 10, 15). Monolingual retrieval performs poorly (Recall@15 <10\%), whereas all CLIR approaches substantially improve retrieval accuracy. Among them, BGE-M3 achieves the highest Recall@15, reaching 96.2\% for Sinhala-English and 95.6\% for Tamil-English, outperforming the best QT approach (Google Translate: 92.4\% and 93.0\%) while avoiding translation overhead. These results demonstrate that multilingual embedding models provide a more effective and scalable solution for cross-lingual retrieval-augmented generation (RAG) in low-resource government domains.

\end{abstract}


\section{Introduction}
Retrieval-Augmented Generation (RAG) improves the factual grounding of large language models by retrieving relevant external content at inference time \cite{lewis2020rag,gao2023rag}. Most RAG systems assume the query and document languages coincide \cite{lewis2020rag}, which limits their usefulness for speakers of low-resource languages who wish to query in their native language while retrieving from a predominantly English-language corpus. This mismatch is acute in Sri Lanka, where citizens commonly interact with public digital services in Sinhala or Tamil while official information remains largely English-centric.

Existing retrieval models are developed and optimized chiefly for high-resource languages, leaving Sinhala and Tamil underrepresented due to limited curated datasets, inconsistent domain-specific translation quality, and weak cross-lingual embedding alignment. This work systematically evaluates practical cross-lingual retrieval pipelines for these two languages against a real-world, government-domain English knowledge base.

\subsection{Research Questions}
\begin{itemize}
\item \textbf{RQ1:} How effective is monolingual English retrieval when queries are issued in Sinhala and Tamil without cross-lingual handling?
\item \textbf{RQ2:} How do query translation-based retrieval approaches compare against cross-lingual embedding-based approaches for Sinhala--English and Tamil--English retrieval?
\item \textbf{RQ3:} Which cross-lingual retrieval strategy is most effective and robust for low-resource Sri Lankan languages in a government-domain setting?
\end{itemize}

\subsection{Contributions}
This study contributes: (i) a systematic comparison of monolingual, query-translation, and cross-lingual-embedding retrieval pipelines for Sinhala--English and Tamil--English retrieval; (ii) a retrieval-ready English knowledge base built from real Sri Lankan government service pages, segmented into 1,699 semantically coherent contexts; (iii) empirical evidence that cross-lingual embeddings, particularly BGE-M3, can outperform translation-based retrieval while removing the translation step entirely; and (iv) a verified, publicly available multilingual (English--Sinhala--Tamil) question-answer benchmark for future CLIR research.

\section{Related Work}
Traditional RAG assumes alignment between query and document language \cite{lewis2020rag}, restricting applicability for low-resource language speakers \cite{shi2021crosslingual}. Retrieval must bridge this linguistic gap by locating passages that are semantically equivalent to the query despite being in a different language \cite{goworek2025bridging}; a weak retrieval component propagates errors into hallucinated generation \cite{saleh2020document,huang2021mixed}.

Monolingual embeddings optimized for a single language leave semantically equivalent cross-lingual pairs far apart in vector space \cite{feng2020labse,shi2021crosslingual}, motivating two established CLIR strategies. Query translation converts the query into the document language before retrieval but is sensitive to translation ambiguity, especially for short, under-specified queries \cite{saleh2020document,huang2021mixed}. Document translation avoids this ambiguity but introduces substantial latency and translation noise, making it impractical at scale \cite{shi2021crosslingual}.

Cross-lingual embeddings instead learn a shared representation space across languages, making semantic similarity language-agnostic \cite{feng2020labse}. Early models such as mBERT \cite{feng2020labse} and XLM-R \cite{conneau2019unsupervised} targeted general multilingual alignment, while LaBSE \cite{feng2020labse} and multilingual E5 \cite{wang2024multilingual} focused on stronger sentence-level cross-lingual retrieval, and BGE-M3 \cite{chen2024bge} extended this with multi-functional dense and hybrid retrieval training. Translation-based pipelines using NLLB \cite{nllb2022} or high-quality commercial MT remain competitive baselines, but recent multilingual embeddings are closing, and in some cases exceeding, this gap \cite{goworek2025bridging}. General-purpose multilingual retrieval benchmarks such as MIRACL \cite{zhang2023miracl} and mMARCO \cite{bonifacio2021mmarco} cover a broad set of languages but do not include Sinhala or Tamil and are not grounded in a single, government-domain corpus, which limits their ability to characterize retrieval behavior for these two languages in a realistic, terminology-heavy public-service setting. Prior work has not systematically benchmarked government-domain CLIR for Sinhala and Tamil, or directly compared QT and CLE pipelines for these two typologically distinct, low-resource South Asian languages, which this study addresses \cite{zhang2021mind}.

\section{Methodology}
This section describes the proposed comparative retrieval architecture used to evaluate cross-lingual information access for Sinhala--English and Tamil--English queries.

\subsection{Overview of the Proposed Architecture}
\begin{figure*}[t]
\centering
\begin{tikzpicture}[
  node distance=6mm and 12mm,
  box/.style={rectangle, draw, rounded corners, align=center, fill=gray!8,
              minimum width=3.0cm, minimum height=0.9cm, font=\small},
  lbl/.style={font=\small\bfseries},
  arr/.style={-{Latex[length=2.2mm]}, thick}
]

\node[box] (b1) {Query \textless SI/TA\textgreater};
\node[box, below=of b1] (b2) {English Embeddings\\\textless FastEmbed\textgreater};
\node[box, below=of b2] (b3) {Top-15 Contexts};
\draw[arr] (b1) -- (b2);
\draw[arr] (b2) -- (b3);
\node[lbl, above=3mm of b1] {Baseline RAG};

\node[box, right=of b1] (q1) {Query \textless SI/TA\textgreater};
\node[box, below=of q1] (q2) {Translation Model\\\textless Google Translate /\\ mBART50 / NLLB\textgreater};
\node[box, below=of q2] (q3) {English Embeddings\\\textless FastEmbed\textgreater};
\node[box, below=of q3] (q4) {Top-15 Contexts};
\draw[arr] (q1) -- (q2);
\draw[arr] (q2) -- (q3);
\draw[arr] (q3) -- (q4);
\node[lbl, above=3mm of q1] {QT-based RAG};

\node[box, right=of q1] (c1) {Query \textless SI/TA\textgreater};
\node[box, below=of c1] (c2) {Cross-lingual Embeddings\\\textless LaBSE / E5-M /\\ BGE-M3\textgreater};
\node[box, below=of c2, yshift=-9mm] (c3) {Top-15 Contexts};
\draw[arr] (c1) -- (c2);
\draw[arr] (c2) -- (c3);
\node[lbl, above=3mm of c1] {CLE-based RAG};

\end{tikzpicture}
\caption{Proposed three-pipeline retrieval architecture. All pipelines share the same indexed English knowledge base and return the top-15 ranked contexts by cosine similarity; they differ only in how the Sinhala/Tamil query is bridged into the English embedding space before ranking.}
\label{fig:architecture}
\end{figure*}
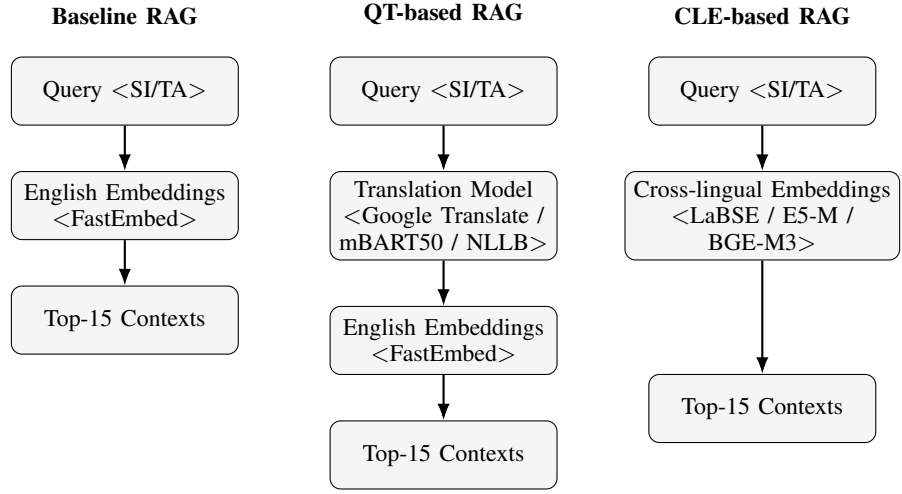

The proposed solution is a controlled, three-pipeline comparative architecture that isolates the effect of cross-lingual adaptation strategy while holding the underlying document index, query set, and evaluation protocol fixed. Each pipeline accepts the same Sinhala or Tamil query and returns a ranked list of the top-15 English contexts from a common indexed knowledge base, differing only in \emph{how} (or whether) the linguistic gap between the query and the English documents is bridged before ranking:
\begin{enumerate}
\item \textbf{Baseline (Monolingual English Embeddings, No Translation)}: the naive cross-lingual case, i.e., no adaptation is applied.
\item \textbf{Query Translation (QT)}: an explicit, translation-based strategy that lexically bridges the language gap using machine translation prior to embedding.
\item \textbf{Cross-Lingual Embedding (CLE)}: an implicit, embedding-based strategy that achieves direct semantic alignment without any translation step.
\end{enumerate}
All three pipelines share a common English document index, built once over the full context collection, and are evaluated independently under identical retrieval-depth and similarity-ranking conditions, isolating retrieval-strategy effects from indexing effects. Fig.~\ref{fig:architecture} illustrates the resulting three-pipeline architecture.

\subsection{Baseline Pipeline}
The baseline pipeline represents the naive cross-lingual case, where no translation or multilingual alignment is applied. The original Sinhala or Tamil query is embedded directly using the same monolingual English embedding model (FastEmbed) that was used to build the document index, and retrieval is performed by cosine similarity between this mismatched, non-English query embedding and the English document vectors. Because the embedding model was never optimized to align Sinhala/Tamil and English representations, this pipeline serves as a lower-bound reference against which the benefit of explicit cross-lingual adaptation can be measured.

\subsection{Query Translation (QT) Pipeline}
The QT pipeline models the explicit translation strategy: the Sinhala or Tamil query is first machine-translated into English, and the translated query is then embedded using the same monolingual English embedding model (FastEmbed) used for the document index, before cosine similarity ranks the English passages. Three translation models, spanning commercial and open-source, and modern and earlier-generation systems, are evaluated within this pipeline:
\begin{itemize}
\item \textbf{Google Translate}: a commercial system with mature infrastructure and high translation quality for low-resource languages, included to provide a practical, near upper-bound QT baseline.
\item \textbf{NLLB (No Language Left Behind)}: an open-source neural machine translation model optimized for low-resource languages, included to assess whether open-source MT can approach commercial translation quality for Sinhala and Tamil.
\item \textbf{mBART50}: a widely cited multilingual sequence-to-sequence model, included to represent an academic MT baseline and to characterize the limitations of earlier multilingual translation architectures when applied to retrieval-focused tasks.
\end{itemize}

\subsection{Cross-Lingual Embedding (CLE) Pipeline}
The CLE pipeline models the implicit, semantic-alignment strategy: the original Sinhala or Tamil query is embedded directly in a multilingual embedding space, and passage ranking is computed via cosine similarity between this non-English query vector and the English document vectors within the same shared space, without any translation step. Three multilingual bi-encoder models are evaluated within this pipeline:
\begin{itemize}
\item \textbf{LaBSE (Language-Agnostic BERT Sentence Embeddings)}: a widely adopted baseline for cross-lingual sentence similarity and retrieval, included as a stable point of reference.
\item \textbf{Multilingual E5}: an instruction-tuned embedding model optimized for retrieval tasks, included to examine the benefit of task-aware, retrieval-oriented multilingual embeddings.
\item \textbf{BGE-M3}: a state-of-the-art multilingual embedding model supporting dense and hybrid retrieval, included for its strong performance on multilingual retrieval benchmarks and its ability to operate in a fully language-agnostic manner without reliance on translation.
\end{itemize}

\subsection{Embedding, Indexing, and Ranking}
Document contexts are vectorized once per embedding model (FastEmbed for the Baseline/QT pipelines; LaBSE, multilingual E5, and BGE-M3 for the CLE pipeline) and stored in separate Pinecone vector stores, in conjunction with LangChain utilities for embedding generation, vector indexing, and similarity-based retrieval. For every pipeline and query, ranking is computed using cosine similarity between the query embedding and the indexed English context embeddings, and results are collected up to a fixed depth of 15 contexts to support Recall@k evaluation across multiple retrieval depths.

\subsection{Implementation Details}
The proposed pipelines were implemented in Python within a Jupyter Notebook environment. Structured data was managed with CSV files and manipulated using pandas and NumPy. GIC service pages were programmatically scraped using BeautifulSoup4, complemented by LangChain-GenAI for automated extraction and organization of textual content, and the resulting embeddings were indexed and queried through Pinecone. Together, these tools enabled a consistent, reproducible implementation of the Baseline, QT, and CLE pipelines described above.

\section{Experiment}

\subsection{Experimental Setup}

\subsubsection{Dataset}
The knowledge base was constructed from Sri Lanka's Government Information Center (GIC) website\footnote{https://gic.gov.lk/}. A total of 761 web pages spanning ten main service categories (e.g., Trade, Agriculture, Health, Education, Justice, Banking) and 70 subcategories were scraped, then segmented via an LLM-guided prompt into 1,699 semantically coherent contexts of 250--400 words each, preserving original wording. A stratified sample of 500 contexts, proportional to category distribution, was selected for evaluation (Table~\ref{tab:dataset}). For each sampled context, an LLM generated a question--answer pair in English, Sinhala, and Tamil, which was subsequently human-verified for linguistic correctness and alignment, yielding 500 verified Sinhala and 500 verified Tamil queries, each mapped to a single gold English context.

\begin{table}[htbp]
\caption{Stratified Context Sample by Service Category}
\begin{center}
\begin{tabular}{|l|c|c|}
\hline
\textbf{Main Category} & \textbf{Original} & \textbf{Sampled} \\
\hline
Trade, Business \& Industry & 419 & 124 \\
\hline
Agriculture, Livestock \& Fisheries & 197 & 59 \\
\hline
Health, Well-being \& Social Service & 168 & 49 \\
\hline
Justice, Law \& Rights & 147 & 43 \\
\hline
Banking, Tax \& Insurance & 113 & 33 \\
\hline
Education \& Training & 110 & 32 \\
\hline
Employment Information & 107 & 31 \\
\hline
Housing, Property \& Utilities & 92 & 27 \\
\hline
Citizen's Registrations & 89 & 26 \\
\hline
Travel, Tourism \& Leisure & 88 & 26 \\
\hline
Environment & 88 & 26 \\
\hline
Communication \& Media & 81 & 24 \\
\hline
\textbf{Total} & \textbf{1699} & \textbf{500} \\
\hline
\end{tabular}
\label{tab:dataset}
\end{center}
\end{table}

All three pipelines described in Section~II (Baseline, QT, CLE) were evaluated under identical conditions using the same 1,699 indexed contexts and 500 Sinhala/Tamil query sets, retrieving up to 15 contexts per query.

\subsubsection{Evaluation Metric}
Because each query maps to exactly one gold English passage, Recall@k was adopted as the primary metric, defined for a query $q$ as 1 if the gold context appears in the top-$k$ retrieved results and 0 otherwise, averaged over all $N$ queries:
\begin{equation}
R@k = \frac{1}{N}\sum_{i=1}^{N} R@k(q_i)
\label{eq:recall}
\end{equation}
Recall was computed at $k = 1, 3, 5, 10,$ and $15$ to capture both top-rank precision and overall retrieval coverage, which is directly relevant to downstream RAG generation quality.

\subsection{Results}
Table~\ref{tab:sinhala} and Table~\ref{tab:tamil} report Recall@k for Sinhala--English and Tamil--English queries, respectively. The baseline pipeline performed extremely poorly for both languages (Recall@15 of 8.2\% for Sinhala and 4.2\% for Tamil), confirming that monolingual English embeddings cannot bridge the linguistic gap (RQ1). All QT and CLE pipelines improved recall substantially over this baseline.

\subsubsection{Why Baseline Retrieval Fails}
This near-zero recall stems from a fundamental misalignment between the linguistic spaces of the query and the document index rather than from any weakness in the retrieval procedure itself. Because the baseline pipeline embeds the Sinhala or Tamil query with a monolingual English embedding model (FastEmbed), it is, in effect, asking a model that has only ever learned an English semantic space to place a non-English query within that space. FastEmbed was optimized for a single language, so semantically equivalent terms across Sinhala/Tamil and English are not co-located in the resulting vector space, i.e., a query and its correct English passage can be conceptually identical while being geometrically distant as embeddings. Cosine similarity ranking is only meaningful when queries and documents are embedded in a shared, aligned space, and no such alignment exists for this pipeline. Consequently, the ranked results returned by the baseline are effectively unrelated to the true semantic content of the query, which explains why recall remains below 10\% at every retrieval depth and why explicit translation or cross-lingual embedding alignment, rather than a larger monolingual index or a deeper retrieval depth, is required to make Sinhala- and Tamil-language retrieval viable.

\begin{table}[htbp]
\caption{Recall@k for Sinhala--English Queries (\%)}
\begin{center}
\begin{tabular}{|l|c|c|c|c|c|}
\hline
\textbf{Approach} & \textbf{R@1} & \textbf{R@3} & \textbf{R@5} & \textbf{R@10} & \textbf{R@15} \\
\hline
CLE - BGE-M3 & 59.4 & 83.4 & 90.2 & 94.0 & \textbf{96.2} \\
\hline
QT - Google Translate & 60.0 & 79.6 & 85.4 & 90.4 & 92.4 \\
\hline
QT - NLLB & 53.8 & 71.4 & 77.6 & 85.4 & 87.0 \\
\hline
QT - mBART50 & 46.2 & 66.8 & 73.8 & 80.0 & 82.8 \\
\hline
CLE - E5 Multilingual & 36.4 & 54.2 & 59.6 & 68.4 & 73.0 \\
\hline
CLE - LaBSE & 30.4 & 49.4 & 59.8 & 68.0 & 72.4 \\
\hline
Baseline (no adaptation) & 1.6 & 3.8 & 5.6 & 7.4 & 8.2 \\
\hline
\end{tabular}
\label{tab:sinhala}
\end{center}
\end{table}

\begin{table}[htbp]
\caption{Recall@k for Tamil--English Queries (\%)}
\begin{center}
\begin{tabular}{|l|c|c|c|c|c|}
\hline
\textbf{Approach} & \textbf{R@1} & \textbf{R@3} & \textbf{R@5} & \textbf{R@10} & \textbf{R@15} \\
\hline
CLE - BGE-M3 & 61.0 & 81.0 & 89.6 & 94.2 & \textbf{95.6} \\
\hline
QT - Google Translate & 59.4 & 78.4 & 85.2 & 91.2 & 93.0 \\
\hline
QT - NLLB & 53.2 & 74.6 & 80.0 & 86.6 & 89.0 \\
\hline
CLE - E5 Multilingual & 48.2 & 71.2 & 78.6 & 85.8 & 88.6 \\
\hline
QT - mBART50 & 48.4 & 68.0 & 76.0 & 82.6 & 85.2 \\
\hline
CLE - LaBSE & 29.0 & 46.2 & 55.4 & 66.8 & 72.4 \\
\hline
Baseline (no adaptation) & 0.6 & 1.8 & 2.0 & 3.8 & 4.2 \\
\hline
\end{tabular}
\label{tab:tamil}
\end{center}
\end{table}

\begin{figure}[htbp]
\centerline{\includegraphics[width=0.95\linewidth]{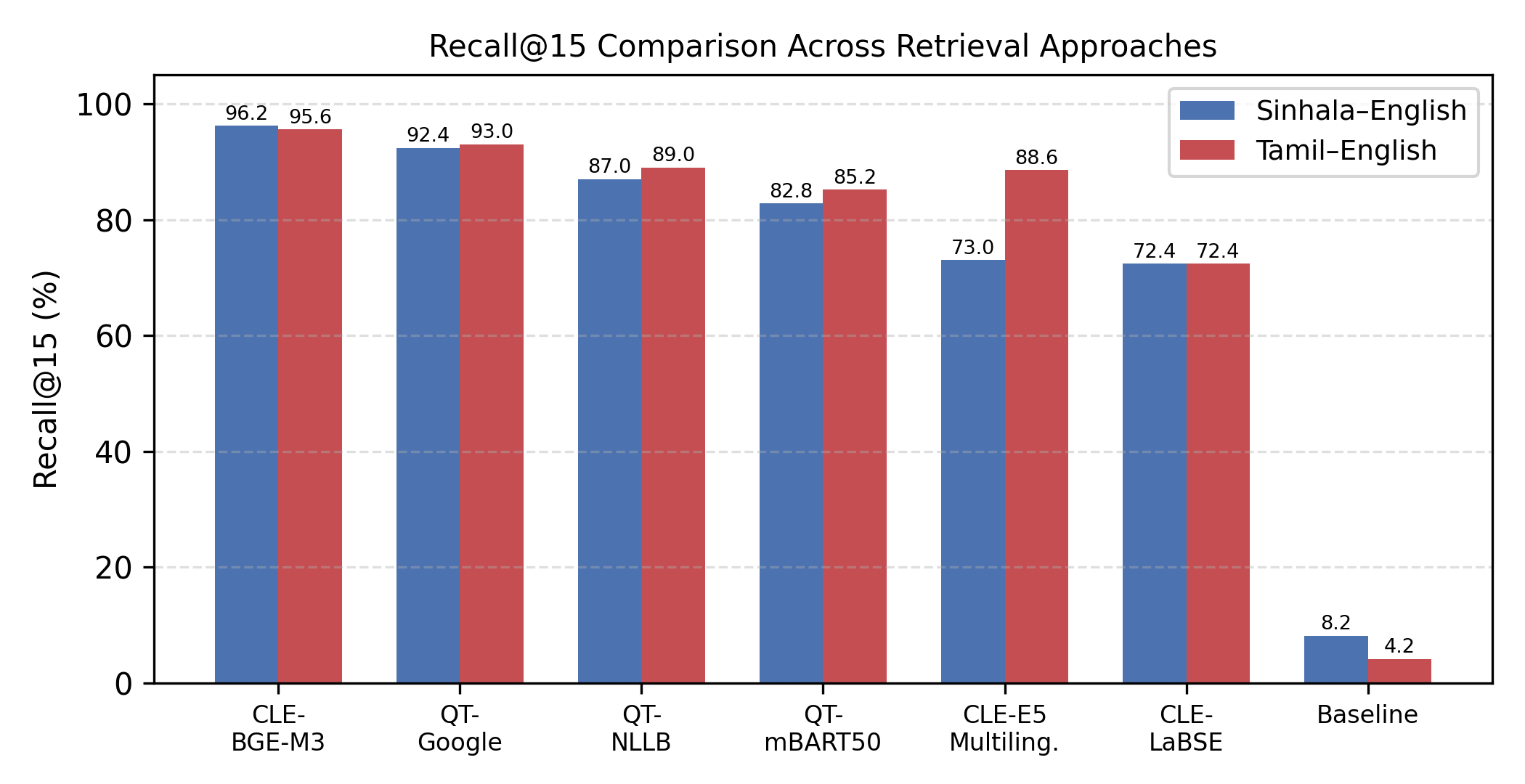}}
\caption{Recall@15 across all evaluated retrieval approaches for Sinhala--English and Tamil--English queries.}
\label{fig:recall15}
\end{figure}

CLE--BGE-M3 achieved the highest Recall@15 for both language pairs (96.2\% Sinhala, 95.6\% Tamil), followed by QT--Google Translate (92.4\%/93.0\%). NLLB (87.0\%/89.0\%) and mBART50 (82.8\%/85.2\%) trailed the commercial translation engine, while CLE--E5 Multilingual (73.0\%/88.6\%) and CLE--LaBSE (72.4\%/72.4\%) trailed BGE-M3 (RQ2), as summarized in Fig.~\ref{fig:recall15}.

\subsection{Ablation Study}
To better understand \emph{why} the pipelines in Section~IV-B differ, this subsection breaks the results down along two axes: the retrieval paradigm (QT vs. CLE) and the individual embedding/translation model used within each paradigm.

\subsubsection{Query Translation vs. Cross-Lingual Embeddings}
\begin{figure}[htbp]
\centerline{\includegraphics[width=0.95\linewidth]{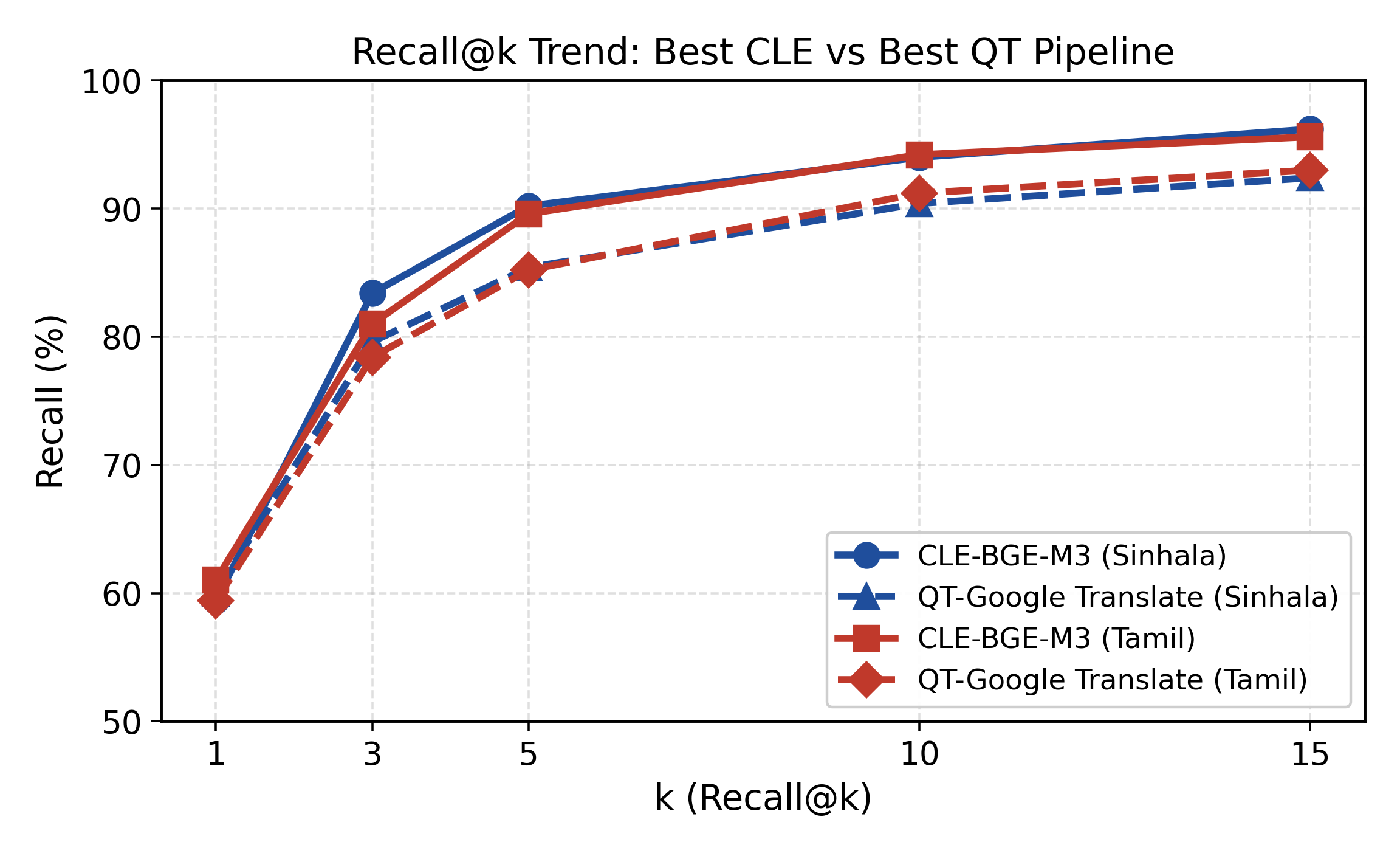}}
\caption{Recall@k trend for the strongest CLE and QT pipelines at increasing retrieval depth.}
\label{fig:trend}
\end{figure}

Google Translate provided a strong QT baseline due to its mature infrastructure and extensive multilingual training data, but open-source alternatives NLLB and mBART50 lagged, likely reflecting weaker handling of the morphological complexity of Sinhala and Tamil and consequent error propagation into retrieval. CLE pipelines avoid this dependency entirely: by mapping queries and documents into a shared semantic space, BGE-M3 is not exposed to lexical translation errors, and it achieved the highest Recall@1 for both languages (approximately 60\%, Fig.~\ref{fig:trend}), indicating that it also ranks the correct passage earlier, which is a property that is particularly valuable in RAG settings where only the top few retrieved passages are used for generation.

\subsubsection{Language-Specific and Model-Level Behavior}
Model performance was not uniform across languages. Table~\ref{tab:gap} quantifies this by reporting the Recall@15 gap ($\Delta$ = Tamil $-$ Sinhala) for every model. Multilingual E5 shows by far the largest gap ($+$15.6 points: 88.6\% Tamil--English vs. 73.0\% Sinhala--English), evidencing that its cross-lingual robustness depends heavily on per-language training-data coverage. A plausible explanation, which this study does not verify directly, is that Tamil is more heavily represented than Sinhala in the large-scale multilingual web corpora (e.g., mC4, CC-100) typically used to pretrain models such as multilingual E5, so its embedding space is likely to be better aligned with English for Tamil than for Sinhala; confirming this would require inspecting E5's pretraining data composition or corpus-size statistics directly, which falls outside the scope of this study. LaBSE, in contrast, shows a gap of exactly 0.0 points (72.4\% for both languages), i.e., the most consistent of any model, although at the lowest overall recall, evidencing its general-purpose (rather than retrieval- or domain-optimized) training objective. BGE-M3 shows a near-zero gap ($-$0.6 points: 96.2\% Sinhala vs. 95.6\% Tamil) while simultaneously achieving the highest recall for both languages, evidencing that its multi-functional dense-and-hybrid retrieval training generalizes across typologically different low-resource languages more effectively than the other CLE models (RQ3). The QT pipelines show smaller, comparatively uniform gaps ($+$0.6 to $+$2.4 points), consistent with translation engines that are not language-pair-specific in the way individual embedding models are.

\begin{table}[htbp]
\caption{Recall@15 Gap Between Tamil and Sinhala by Model}
\begin{center}
\begin{tabular}{|l|c|c|c|}
\hline
\textbf{Approach} & \textbf{Sinhala} & \textbf{Tamil} & \textbf{$\Delta$ (Ta$-$Si)} \\
\hline
CLE - BGE-M3 & 96.2 & 95.6 & $-$0.6 \\
\hline
QT - Google Translate & 92.4 & 93.0 & $+$0.6 \\
\hline
QT - NLLB & 87.0 & 89.0 & $+$2.0 \\
\hline
QT - mBART50 & 82.8 & 85.2 & $+$2.4 \\
\hline
CLE - E5 Multilingual & 73.0 & 88.6 & $+$15.6 \\
\hline
CLE - LaBSE & 72.4 & 72.4 & 0.0 \\
\hline
\end{tabular}
\label{tab:gap}
\end{center}
\end{table}

\section{Conclusion}
This study proposed and evaluated a controlled, three-pipeline retrieval architecture, namely a monolingual baseline, query translation, and cross-lingual embeddings, for Sinhala--English and Tamil--English information access against a real-world Sri Lankan government-domain knowledge base. Monolingual retrieval fails almost entirely for both languages, while all cross-lingual pipelines yield large improvements. BGE-M3 consistently achieved the best and most language-consistent recall (95--96\% at Recall@15) for both languages, outperforming the strongest translation-based pipeline while removing the need for an explicit translation step, and Google Translate remained the strongest QT option with NLLB as a viable open-source alternative. These results position cross-lingual embeddings, particularly BGE-M3, as the more effective and scalable foundation for cross-lingual RAG in low-resource, government-domain settings. These conclusions are based on retrieval-level Recall@k over a single government-services domain, and the evaluated embedding and translation models were used in their pre-trained form without fine-tuning; the observed recall differences are also reported as raw percentages rather than as statistically tested effects. Future work should therefore evaluate end-to-end generation quality, extend the evaluation to additional domains, and explore fine-tuning and hybrid QT/CLE retrieval architectures.

\bibliographystyle{IEEEtran}
\bibliography{references}

\end{document}